\documentclass[english,prb,aps,floatfix,superscriptaddress,twocolumn,longbibliography]{revtex4-1}
\usepackage[T1]{fontenc}
\usepackage[latin9]{inputenc}
\usepackage{color}
\usepackage{amsmath}
\usepackage{amssymb}
\usepackage{graphicx}

\makeatletter

\usepackage{amssymb}
\usepackage{amsmath}
\usepackage{graphicx}
\usepackage[colorlinks=true,linkcolor=blue,anchorcolor=red,citecolor=blue,urlcolor=blue]{hyperref}
\usepackage[caption=false]{subfig}

\makeatother

\usepackage{babel}
\begin{document}
\renewcommand{\figurename}{Fig.}
\title{Influence of a chiral chemical potential on Weyl hybrid junctions}
\author{Daniel Breunig}
\affiliation{Theoretische Physik IV, Institut f\"ur Theoretische Physik und Astrophysik,
Universit\"at W\"urzburg, D-97074 W\"urzburg, Germany}
\author{Song-Bo Zhang}
\email{song-bo.zhang@physik.uni-wuerzburg.de}

\affiliation{Theoretische Physik IV, Institut f\"ur Theoretische Physik und Astrophysik,
Universit\"at W\"urzburg, D-97074 W\"urzburg, Germany}
\author{Martin Stehno}
\affiliation{Experimentelle Physik III, Physikalisches Institut, Universit\"at
W\"urzburg, D-97074 W\"urzburg, Germany}
\author{Bj\"orn Trauzettel}
\affiliation{Theoretische Physik IV, Institut f\"ur Theoretische Physik und Astrophysik,
Universit\"at W\"urzburg, D-97074 W\"urzburg, Germany}
 \affiliation{W\"urzburg-Dresden Cluster of Excellence ct.qmat, Germany}
\date{\today}
\begin{abstract}
We study the transport properties and superconducting proximity effect
in NSN junctions formed by a time-reversal symmetry broken Weyl semimetal
(WSM) in proximity to an $s$-wave superconductor. We find that the
differential conductances and induced pairing amplitudes strongly
depend on the angle between the junction direction in real space and
the axis separating the Weyl nodes in momentum space. We identify
the influence of a chiral chemical potential, i.e., the electron population
imbalance between Weyl nodes of opposite chirality, on the transport
characteristics of the junction. Remarkably, we observe a net spin
polarization of Cooper pairs that are generated via Andreev reflection
in the two WSM regions. The spin polarization is opposite in the two
WSM regions and highly sensitive to the chirality imbalance and excitation
energy.
\end{abstract}
\maketitle

\section{Introduction}

Weyl semimetals (WSMs), which are three-dimensional topological phases
of matter with strong spin-orbit coupling, have been subject of intense
research activities during the last years \citep{Hosur13Physique,Turner2013arXiv,Bernevig15natphy,Burkov2016Nmat,Armitage18RMP}.
They are featured by pairs of linear band crossings, called Weyl nodes,
in momentum space. Since the Weyl nodes are topologically protected
and closely associated with the chiral anomaly, WSMs impose a plethora
of characteristic phenomena, such as surface Fermi arcs \citep{Wan11prb},
nonlocal transport \citep{Parameswaran14prx}, and anomalous magnetoconductance
\citep{Nielsen83plb,Zyuzin12prb,Son13prb,Burkov14prl-chiral,Lu15Weyl-shortrange,ZhangSB16NJP}.
Moreover, there have been a growing number of realistic materials
proposed theoretically \citep{Burkov11prl,Halasz12PRB,Hirayama14PRL,Weng15prx,Huang15nc,Rauch15PRL,Ruan16Ncomms,Ruan16PRL,ZJWang16PRL,YJin17PRB,HYang17NJP,GQChang18PRB}
and confirmed experimentally \citep{Xu15sci-TaAs,Yang15Nphys,Lv15prx,Lv15nphys,Xu15np-NbAs,XuN16NC-TaP}
as WSMs. The proximity of materials with strong spin-orbit coupling
to an $s$-wave superconductor can produce spin-triplet pairing which
may give rise to topological superconductivity \citep{Fu08PRL}. This
implies intriguing physics, such as the emergence of odd-frequency
superconductivity \citep{Burset15PRB,Sato17RPP,Fleckenstein18PRB}
and Majorana bound states \citep{Fu08PRL}. Among spin-triplet pairing,
equal-spin pairing is of particular interest for superconducting spintronics
\citep{linder15nphys} because it offers the potential to transfer
and manipulate spin-polarized Cooper pairs. Recently, it has been
put forward to generate equal-spin pairing in surface states of topological
insulators without introducing magnetic order \citep{Breunig18PRL}.

WSMs possess desirably strong spin-orbit coupling\ \citep{Wan11prb,Armitage18RMP}.
It is therefore of fundamental interest and application potential
for spintronics to explore the possibility of equal-spin pairing and
related transport signatures in WSMs. There have been a few works
studying the transport and pairing properties of superconducting heterostructures
based on WSMs \citep{WChen13EPL,Uchida14JPSJ,Khanna14PRB,Bovenzi17PRB,Chen17PRB,Songbo18PRB,Madsen17PRB,ZhangSB18PRL,Alidoust18PRB,Uddin19PRB}.
However, the induced pairing in the WSM region is still poorly understood.
Furthermore, previous studies focused only on the case where the electron
populations at any Weyl nodes are the same. In principle, Weyl nodes
of opposite chirality do not need to have the same electron population.
In fact, an electron population imbalance between Weyl nodes of different
chirality, called chiral chemical potential (CCP), can be achieved,
e.g., by the chiral anomaly with applying parallel electric and magnetic
fields \citep{Nielsen83plb,Fukushima08PRD,Li16np}, by a strain deformation
\citep{ZDSong16PRB,Cortijo16PRB} or in a superlattice system with
breaking both time-reversal and inversion symmetries \citep{Zyuzin12PRBb}.
This raises an important question of how the CCP influences the transport
properties and superconducting proximity effect in Weyl heterostructures.

In this article, we investigate the transport and induced local pairing
amplitudes in Weyl semimetal-superconductor-semimetal (NSN) junctions
by using the scattering approach \citep{McMillan68PR,Blonder1982,Kashiwaya00RPP}.
In particular, we consider a WSM which breaks time-reversal symmetry
and focus on the influence of a CCP. Due to the anisotropy of the
band structure of the WSM, the differential conductances across the
junction and the superconducting proximity effect strongly depend
on the angle between the junction direction and the axis separating
the Weyl nodes.  The CCP shifts the excitation energy
spectrum of the system and consequently changes the local and nonlocal
differential conductances as a function of the bias voltage. The
interplay of strong spin-orbit coupling and $s$-wave superconductivity
in the system leads to the emergence of equal-spin pairing even deep
inside the WSM regions. Interestingly, we find that for appropriate
angles, the CCP gives rise to a large net spin polarization of Cooper
pairs in the WSM regions. The spin polarization is opposite in the
two WSM regions. It depends on the excitation energy, strength of
the CCP and junction direction. This CCP-induced spin polarization
of Cooper pairs with a dipole feature may find promising applications
in superconducting spintronics.

The remainder of the paper is structured as follows. In Sec.\ \ref{sec:Setup-and-model},
we introduce the setup and model Hamiltonian. Then, we describe in
Sec.\ \ref{sec:Scattering-approach} the scattering states for calculating
the differential conductances and induced pairing amplitudes in the
WSM regions. Sec.\ \ref{sec:Differential-conductance} is dedicated
to the results regarding the differential conductances, and Sec.\ \ref{sec:Induced-pairing-amplitudes}
to the results regarding pairing amplitudes and spin polarization
of Cooper pairs in the WSM regions. Finally, we conclude and discuss
the experimental relevance in Sec.\ \ref{sec:Discussion-and-conclusion}.

\begin{figure}[h]
\centering

\includegraphics[width=8cm]{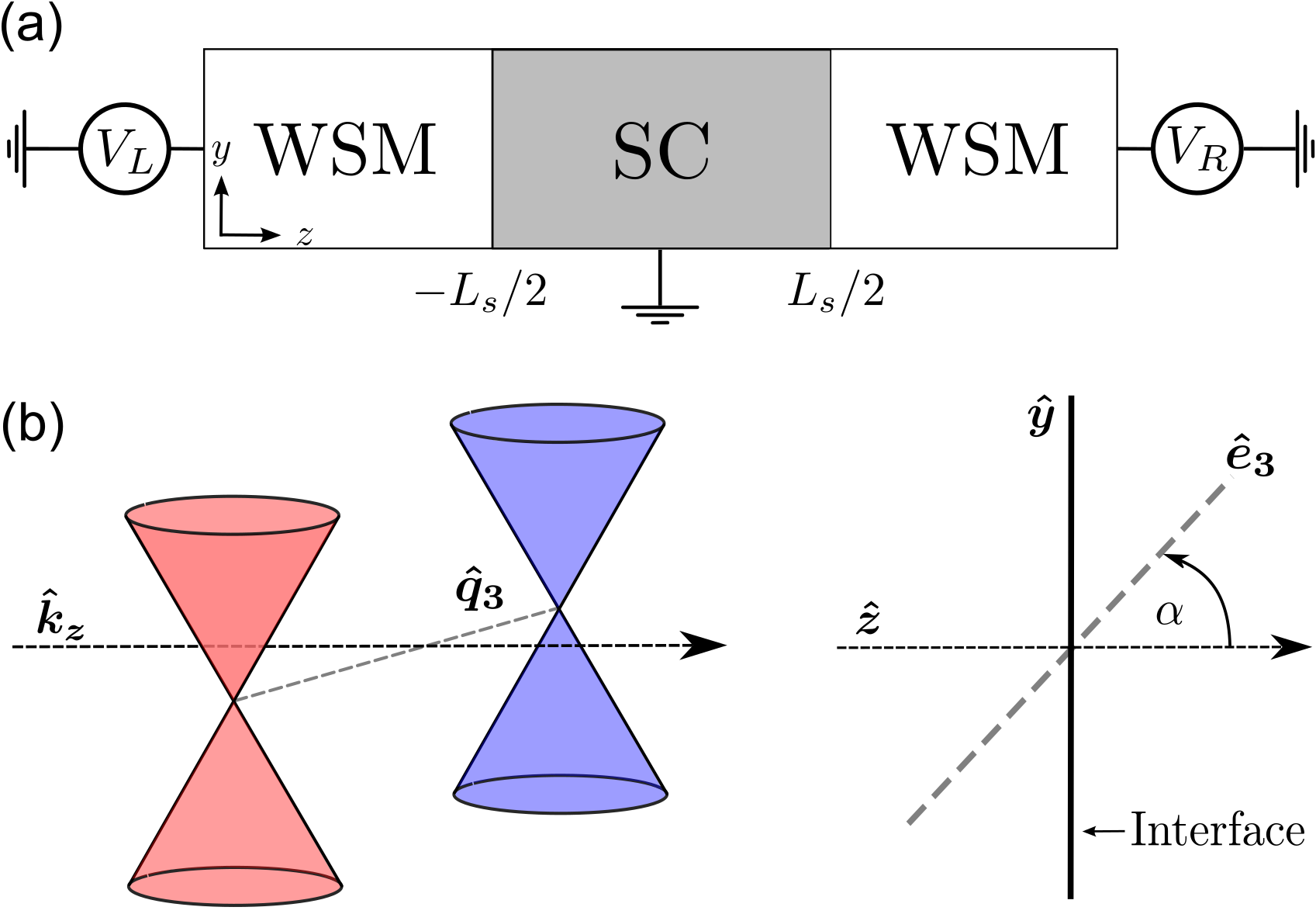}

\caption{(a) An NSN junction formed by a Weyl superconductor sandwiched by
two WSM regions (or leads)  along $\hat{z}$ direction.
The superconductor is grounded while the two WSM regions
are connected to the voltage sources $V_{L}$ and $V_{R}$, respectively.
(b) The junction direction  $\hat{z}$ deviates
from the axis $\hat{e}_{3}$ separating the Weyl nodes by an angle
$\alpha$. The blue and red colored cones, separated in $\hat{q}_{3}$
direction, denote the Weyl fermions of positive and negative chirality,
respectively.}

\label{fig:setup}
\end{figure}

\section{Setup and model\label{sec:Setup-and-model}}

We consider an NSN junction formed by a Weyl superconductor sandwiched
between two leads of WSMs, as illustrated in Fig.\ \ref{fig:setup}(a).
The left and right WSM regions extend semi-infinitely along $\hat{z}$
direction and are connected to the voltage sources $V_{L}$ and $V_{R}$,
respectively. The superconductor is grounded. The interfaces between
WSMs and superconductor, located at $z_{L}\equiv-L_{s}/2$ and $z_{R}\equiv L_{s}/2$,
are assumed to extend along the $xy$\textendash plane. We consider
the simplest WSM with a single pair of Weyl nodes. The minimal model
at low energies is given by
\begin{align}
 & H_{0}=\sum\limits _{\tau=\pm}{\displaystyle \sum_{{\bf q}}}'\Psi_{\tau{\bf q}}^{\dagger}H_{\tau}({\bf q})\Psi_{\tau{\bf q}},\\
 & H_{\pm}({\bf q})=v_{F}[q_{1}s_{1}+q_{2}s_{2}\mp(q_{3}\mp K_{0})s_{3}]-\mu s_{0},\label{eq:Weyl-Hamiltonian}
\end{align}
where $v_{F}$ is the Fermi velocity which we take as $v_{F}\equiv1$;
$\mu$ is the chemical potential; $\tau=\pm$ denotes the Weyl nodes
of positive and negative chirality, respectively; $s_{0}$ and $s_{1,2,3}$
are unit and Pauli matrices acting on the spin space, respectively.
$\Psi_{\tau{\bf q}}^{\dagger}=(c_{\uparrow,\tau,{\bf q}}^{\dagger},c_{\downarrow,\tau,{\bf q}}^{\dagger})$
is the spinor basis for Weyl nodes with $c_{s,\tau,{\bf q}}^{\dagger}$
the creation operator for an electron with spin $s$ and momentum
${\bf q}$ at node $\tau$. $\sum_{{\bf q}}'$ means that ${\bf q}$
is restricted to the momenta in the vicinity of the nodes. The two
Weyl nodes are related by inversion symmetry, as indicated by $s_{z}H_{+}({\bf q})s_{z}=H_{-}(-{\bf q})$.
Without loss of generality, we assume that the pair of Weyl nodes
are separated in the $\hat{e}_{3}$ crystalline axis, ${\bf K}_{0}=\pm K_{0}\hat{q}_{3}$.
The model\ (\ref{eq:Weyl-Hamiltonian}) also features a rotational
symmetry with respect to the $\hat{e}_{3}$ axis. In general, the
crystal coordinates are different from the junction coordinates. To
describe this, we introduce the angle $\alpha$ between the junction
direction $\hat{z}$ and the $\hat{e}_{3}$ axis. For the junction
problems, it is convenient to work with the junction coordinates ${\bf r}=(x,y,z)$
(correspondingly, ${\bf k}=(k_{x},k_{y},k_{z})$ in momentum space).
The two coordinate systems are related by a rotation matrix with respect
to $\hat{y}$ direction, which is given by
\begin{align}
\hat{R}= & \begin{pmatrix}\cos\alpha & 0 & -\sin\alpha\\
0 & 1 & 0\\
\sin\alpha & 0 & \cos\alpha
\end{pmatrix}.\label{eq:Rotation-matrix}
\end{align}
Explicitly, we use the rotation $(q_{x},q_{y},q_{z})^{T}=\hat{R}(k_{x},k_{y},k_{z})^{T}$
and $(s_{1},s_{2},s_{3})^{T}=\hat{R}(\sigma_{x},\sigma_{y},\sigma_{z})^{T}$
with $T$ meaning the transpose.

We consider Bardeen-Cooper-Schrieffer (BCS)-like pairing in the superconductor,
which may be induced by the proximity to a conventional superconductor
\citep{Meng12PRB,Khanna14PRB}. At low energies, the pairing term
reads
\begin{align}
\mathrm{H}_{P}={\displaystyle \sum_{{\bf q}}}'(\Delta\,c_{\uparrow,+,{\bf q}}^{\dagger}c_{\downarrow,-,-{\bf q}}^{\dagger}+h.c. & ),\label{eq:pairing-term}
\end{align}
where $\Delta$ is the pairing potential. The pairing potential couples
excitations at one Weyl node to those at the other node. Combining
the model\ (\ref{eq:Weyl-Hamiltonian}) with the pairing term in
Eq.\ (\ref{eq:pairing-term}), the Bogoliubov-de Gennes (BdG) Hamiltonian
for the system can be written as two decoupled blocks $\mathcal{H}^{\pm}$
in the basis $(c_{\uparrow,+,{\bf k}}^{\dagger},c_{\downarrow,+,{\bf k}}^{\dagger},c_{\downarrow,-,-{\bf k}},-c_{\uparrow,-,-{\bf k}})$
and $(c_{\uparrow,-,{\bf k}}^{\dagger},c_{\downarrow,-,{\bf k}}^{\dagger},c_{\downarrow,+,-{\bf k}},-c_{\uparrow,+,-{\bf k}})$
in the junction coordinates, respectively. The two blocks are related
by particle-hole symmetry. For a uniform pairing potential, the system
realizes a Weyl superconductor \citep{Meng12PRB,Yang14PRL} with four
Weyl nodes in the BdG spectrum. We next perform the unitary transformations
\begin{equation}
\mathcal{H}^{\pm}\to\hat{\mathrm{U}}_{\alpha}^{\pm}\mathcal{H}^{\pm}(\hat{\mathrm{U}}_{\alpha}^{\pm})^{-1}\label{eq:unitary-transformation}
\end{equation}
with
\begin{equation}
\hat{\mathrm{U}}_{\alpha}^{\pm}=\frac{1}{2}\Big[(\tau_{0}\pm\tau_{z})\sigma_{x}e^{i\alpha{\bf \sigma}_{y}}+(\tau_{0}\mp\tau_{z})\sigma_{0}\Big]e^{\pm iK_{0}.r},
\end{equation}
where $\tau_{0}$ and $\tau_{x,y,z}$ are unit and Pauli matrices
in particle-hole space, respectively. Then, the $K_{0}$-dependence
is moved into the basis wavefunctions and the $\alpha$-dependence
into the order parameter. The resulting BdG Hamiltonians $\mathcal{H}^{\pm}$
become
\begin{equation}
\mathcal{H}^{\pm}=\begin{pmatrix}k_{z}-\mu & k_{x}\pm ik_{y} & -\Delta\sin\alpha & \Delta\cos\alpha\\
k_{x}\mp ik_{y} & -k_{z}-\mu & \Delta\cos\alpha & \Delta\sin\alpha\\
-\Delta\sin\alpha & \Delta\cos\alpha & \mu-k_{z} & -k_{x}\pm ik_{y}\\
\Delta\cos\alpha & \Delta\sin\alpha & -k_{x}\mp ik_{y} & \mu+k_{z}
\end{pmatrix}.\label{eq:Block-Ham}
\end{equation}
In the NSN junction, the chemical and pairing potentials are spatially
dependent. For simplicity, we assume a step-like model for these potentials
\begin{align}
\mu(z) & =\mu_{N}\Theta(|z|-L_{s}/2)+\mu_{S}\Theta(L_{s}/2-|z|),\\
\Delta(z) & =\Delta_{0}\Theta(L_{s}/2-|z|),
\end{align}
which is justified when the chemical potential $\mu_{S}$ in the superconductor
is much larger than that in the WSM regions \citep{beenakker92review,Cayssol08PRL}.
Also to satisfy the mean-field approximation for $\Delta_{0}$, a
heavily doped superconductor is assumed. Thus, in this work, we focus
on the case with $|\mu_{S}|\gg|\mu_{N}|,\Delta_{0}$. A CCP between
the two Weyl nodes, which can be induced by applying parallel electric
and magnetic fields via the chiral anomaly \citep{Nielsen83plb,Fukushima08PRD,Li16np}
or by a strain deformation \citep{Cortijo16PRB,ZDSong16PRB}, can
be taken into account by introducing a term $\mp\chi s_{0}$ in $H_{\pm}$
in Eq.\ (\ref{eq:Weyl-Hamiltonian}), respectively, where $\chi$
measures the strength of the chirality imbalance. This CCP breaks
the inversion symmetry of the model and requires an additional term
\begin{equation}
\mathcal{H}_{\chi}^{\pm}=\mp\chi\tau_{0}\sigma_{0}
\end{equation}
in the BdG Hamiltonian $\mathcal{H}^{\pm}$ in Eq.\ (\ref{eq:Block-Ham}).
To illustrate the main effect of the CCP and for simplicity, we consider
a constant $\chi$ everywhere in the junction. Then, the CCP leads
to opposite energy shifts in the excitation spectra of the two blocks
$\mathcal{H}^{\pm}$. This essential mechanism results in interesting
phenomena with respect to transport and induced pairing amplitudes
as we will discuss below.

\section{Scattering states\label{sec:Scattering-approach}}

To study the transport properties and superconducting proximity effect
in the Weyl NSN junction, we apply the scattering approach \citep{Blonder1982,Kashiwaya00RPP,McMillan68PR}.
Due to the translational invariance parallel to the interfaces, the
scattering modes with different transverse wave vectors ${\bf k}_{\parallel}=(k_{x},k_{y})$
can be treated separately. In the following, we take the $\mathcal{H}^{+}$
block to illustrate the calculations. The results for the other block
$\mathcal{H}^{-}$ can be obtained from those for $\mathcal{H}^{+}$
by exploiting the particle-hole symmetry between them. The four scattering
states for $\mathcal{H}^{+}$ are built up as\begin{subequations}\label{sec:Scatteringstate}
\begin{align}
\phi_{1/2}(z) & =\begin{cases}
\psi_{\stackrel{\rightarrow}{e/h}}(z)+a_{1/2}\psi_{\stackrel{\leftarrow}{h/e}}(z)\\
+b_{1/2}\psi_{\stackrel{\leftarrow}{e/h}}(z), & z<z_{L},\\
\sum\limits _{i=1}^{4}s_{1/2,i}\psi_{i}^{S}(z), & |z|<z_{R},\\
c_{1/2}\psi_{\stackrel{\rightarrow}{e/h}}(z)+d_{1/2}\psi_{\stackrel{\rightarrow}{h/e}}(z), & z>z_{R},
\end{cases}\label{eq:ScatteringState1}\\
\phi_{3/4}(z) & =\begin{cases}
c_{3/4}\psi_{\stackrel{\leftarrow}{e/h}}(z)+d_{3/4}\psi_{\stackrel{\leftarrow}{h/e}}(z), & z<z_{L},\\
\sum\limits _{i=1}^{4}s_{3/4,i}\psi_{i}^{S}(z), & |z|<z_{R},\\
\psi_{\stackrel{\leftarrow}{e/h}}(z)+a_{3/4}\psi_{\stackrel{\rightarrow}{h/e}}(z)\\
+b_{3/4}\psi_{\stackrel{\rightarrow}{e/h}}(z), & z>z_{R},
\end{cases}\label{eq:ScatteringState2}
\end{align}
\end{subequations}where $\phi_{1/2}(z)$ describes an electron/hole
excited in the left lead moving towards the interface, while $\phi_{3/4}(z)$
describes the corresponding processes in the right lead. We omit the
factor $e^{ik_{x}x+ik_{y}y}$ for convenience. $a_{j},b_{j},c_{j}$
and $d_{j}$ with $j\in\{1,2,3,4\}$ are amplitudes of Andreev reflection,
normal reflection, electron co-tunneling and crossed Andreev reflection,
respectively, while $s_{j,i}$ with $i\in\{1,2,3,4\}$ are the scattering
amplitudes in the superconducting region. These amplitudes are determined
by matching the wavefunctions at the interfaces $z=z_{L/R}$,
\begin{align}
\phi_{j}(z_{L/R}-0^{+}) & =\phi_{j}(z_{L/R}+0^{+}).\label{eq:match}
\end{align}

The basis wavefunctions in the WSMs are given by\begin{subequations}
\begin{align}
\psi_{\overrightarrow{e}}(z) & =\left(J_{e},k_{||}e^{-i\theta_{k}},0,0\right)^{T}e^{ik_{e}z},\\
\psi_{\overleftarrow{e}}(z) & =\left(k_{||}e^{i\theta_{k}},J_{e},0,0\right)^{T}e^{-ik_{e}z},\\
\psi_{\overrightarrow{h}}(z) & =\left(0,0,k_{||}e^{-i\theta_{k}},-J_{h}\right)^{T}e^{ik_{h}z},\\
\psi_{\overleftarrow{h}}(z) & =\left(0,0,J_{h},-k_{||}e^{i\theta_{k}}\right)^{T}e^{-ik_{h}z},
\end{align}
\end{subequations}where $k_{e/h}=\zeta_{e/h}\sqrt{(\varepsilon+\chi\pm\mu_{N})^{2}-k_{||}^{2}}$,
$J_{e/h}=k_{e/h}+\varepsilon+\chi\pm\mu_{N}$, $k_{||}=|{\bf k}_{\parallel}|$
and $\theta_{k}=\arg\left(k_{y}/k_{x}\right)$. The alphabetical subscript
$e/h$ distinguishes electrons from holes, the arrows point in the
direction of propagation (with respect to $\hat{z}$ direction) and
$\zeta_{e/h}=\text{sign}\left(\varepsilon+\chi\pm\mu_{N}+k_{||}\right)$
indicates whether the particle stems form the valence or the conduction
band.

The basis wavefunctions in the superconductor, in general, can also
be found analytically, but are too extensive to be displayed here.
However, under the assumption of a large chemical potential mismatch,
$|\mu_{S}|\gg|\mu_{N}|$, all excitations show quasi-perpendicular
($k_{||}\approx0$) transmission into the superconductor. In this
regime, we can approximate the basis wavefunctions as\begin{subequations}
\begin{align}
\psi_{1}^{S}(z) & =\left(K_{e},0,-\Delta_{0}\sin\alpha,0\right)^{T}e^{ik_{q}z}e^{i\mu_{S}z},\\
\psi_{2}^{S}(z) & =\left(0,K_{e},0,\Delta_{0}\sin\alpha,\right)^{T}e^{-ik_{q}z}e^{-i\mu_{S}z},\\
\psi_{3}^{S}(z) & =\left(0,\Delta_{0}\sin\alpha,0,K_{e}\right)^{T}e^{ik_{q}z}e^{-i\mu_{S}z},\\
\psi_{4}^{S}(z) & =\left(\Delta_{0}\sin\alpha,0,K_{e},0\right)^{T}e^{-ik_{q}z}e^{i\mu_{S}z},
\end{align}
\end{subequations}where $K_{e}=\varepsilon+\chi+k_{q}$ and $k_{q}=\sqrt{(\varepsilon+\chi)^{2}-\Delta_{0}^{2}\sin^{2}\alpha}$.
Evidently, the effective superconducting gap depends on the angle
$\alpha$ as $\tilde{\Delta}_{0}\equiv\Delta_{0}|\sin\alpha|$. Hence,
it vanishes for integer multiples and is maximal for half integer
multiples of $\pi$. This angle dependence reflects the anisotropy
of the band structure and may be regarded as a characteristic feature
of BCS-like superconductivity in time-reversal symmetry broken Weyl hybrid structures \cite{WChen13EPL,Madsen17PRB,Uddin19PRB}.
\\

\section{Differential conductance\label{sec:Differential-conductance}}

With the scattering amplitudes $a_{j},b_{j},c_{j}$ and $d_{j}$,
we are ready to calculate the differential conductances across the
junction. The local and non-local (differential) conductances, defined
as $G_{\text{LL}}\equiv dI_{L}/dV_{L}|_{V_{R}=0}$ and $G_{\text{LR}}\equiv dI_{R}/dV_{L}|_{V_{R}=0}$,
respectively, at zero temperature can be written in terms of scattering
probabilities\ \citep{Lambert93JPCM,Anantram96PRB}

\begin{widetext}

\begin{figure}[h]
\centering

\captionsetup{justification=raggedright}

\includegraphics[width=17cm]{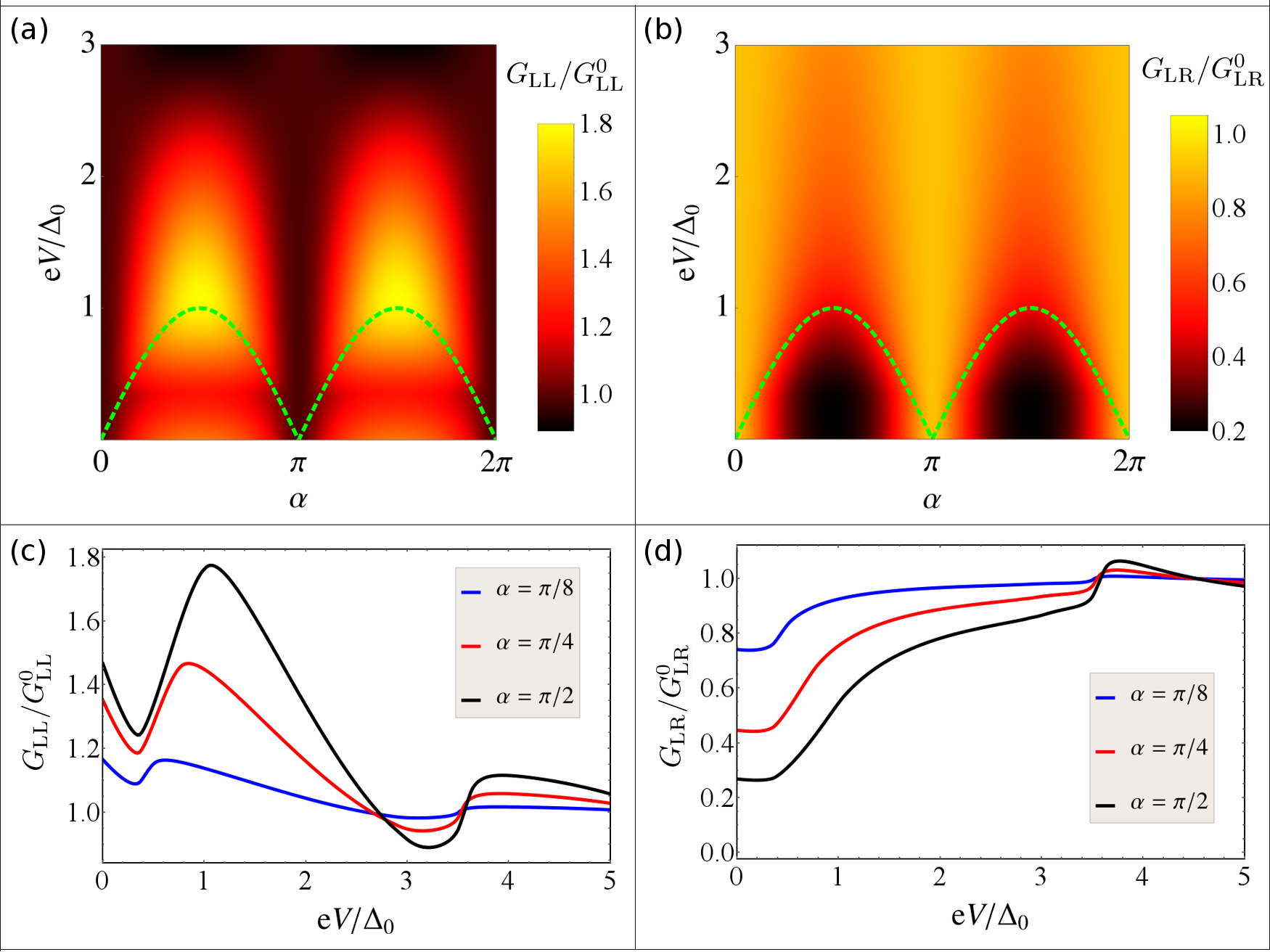}

\caption{Contour plots of the (a) local and (b) non-local conductances $\bar{G}_{\text{L\ensuremath{\Lambda}}}$
with $\text{\ensuremath{\Lambda}}\in\{\text{L,R}\}$ as functions
of the angle $\alpha$ and bias voltage $eV$. The green dashed curves
indicate the effective superconducting gap $\tilde{\Delta}_{0}(\alpha)$.
(c) Local and (d) non-local conductances $\bar{G}_{\text{L\ensuremath{\Lambda}}}$
as functions of $eV$ for different choices of $\alpha$. For different
$\alpha$$(\protect\notin\{0,\pi\})$, the curves of $\bar{G}_{\text{L\ensuremath{\Lambda}}}(eV)$
are qualitatively the same. $\mu_{S}=10^{6}\Delta_{0}$,
$\mu_{N}=10^{3}\Delta_{0}$, $\chi=0$ and $L_{s}=\xi$ for all plots.}

\label{fig:conductance}
\end{figure}

 \end{widetext}
 \begin{subequations}\label{eq:formula-conductance}
\begin{align}
G_{\text{LL}} & =\frac{2\mathrm{e}^{2}}{h}\sum_{{\bf k}_{\parallel}}\left(1+R_{eh}-R_{ee}\right),\label{eq:local-conductance}\\
G_{\text{LR}} & =\frac{2\mathrm{e}^{2}}{h}\sum_{{\bf k}_{\parallel}}\left(T_{ee}-T_{eh}\right),\label{eq:nonlocal-conductance}
\end{align}
\end{subequations}where the sum runs over all relevant ${\bf k}_{\parallel}$
that allow incident channels \footnote{In this work, we consider the transport in the NSN
junction as a two-terminal scattering problem. This is perfectly justified
in the sub-gap regime where Cooper pairs govern the transport in the
superconductor. In experiments, quasi-particles may escape into the
superconducting lead, especially in the supra-gap regime, which may
modify the results quantitatively. However, our treatment should be
valid if the quasi-particle escape is small compared to other transport
processes.}. The probabilities of the four relevant processes for an incident
electron from the left are given by\begin{subequations}
\begin{align}
R_{ee} & =|b_{1}|^{2},\\
R_{eh} & =\frac{J_{h}}{J_{e}}\frac{\mathrm{Re}\left(k_{h}\right)}{k_{e}}|a_{1}|^{2},\\
T_{ee} & =|c_{1}|^{2},\\
T_{eh} & =\frac{J_{h}}{J_{e}}\frac{\mathrm{Re}\left(k_{h}\right)}{k_{e}}|d_{1}|^{2}.
\end{align}
\end{subequations} Let us assume a bias voltage
$V_{L}=V$ is applied to the left lead and keep the right lead unbiased
$V_{R}=0$. The bias voltage $V$ enters the formulas \eqref{eq:local-conductance} and \eqref{eq:nonlocal-conductance}
as the excitation energy $\varepsilon\rightarrow eV$ via the scattering
probabilities. For definiteness and experimental
feasibility, we consider a large chemical potential in the WSMs compared
to the pairing potential and CCP, i.e., $|\mu_{N}|\gg\Delta_{0},\chi$.
To manifest the non-local transport, we choose a junction length comparable
to the coherence length of the superconductor, $L_{s}=\xi\equiv v_{F}/\Delta_{0}$.
We note that the main results discussed below do not change qualitatively
for other choices of $\mu_{N}$ or $L_{s}$. In the following, we
normalize the local and non-local conductances $\bar{G}_{\text{L\ensuremath{\Lambda}}}\equiv G_{\text{L\ensuremath{\Lambda}}}/G_{\text{L\ensuremath{\Lambda}}}^{0}$,
$\text{\ensuremath{\Lambda}}\in\{\text{L,R}\}$ with respect to their
normal counterparts $G_{\text{L\ensuremath{\Lambda}}}^{0}$ at $\Delta_{0}=0$,
respectively.

\begin{figure}[h]
\centering

\includegraphics[width=8.5cm]{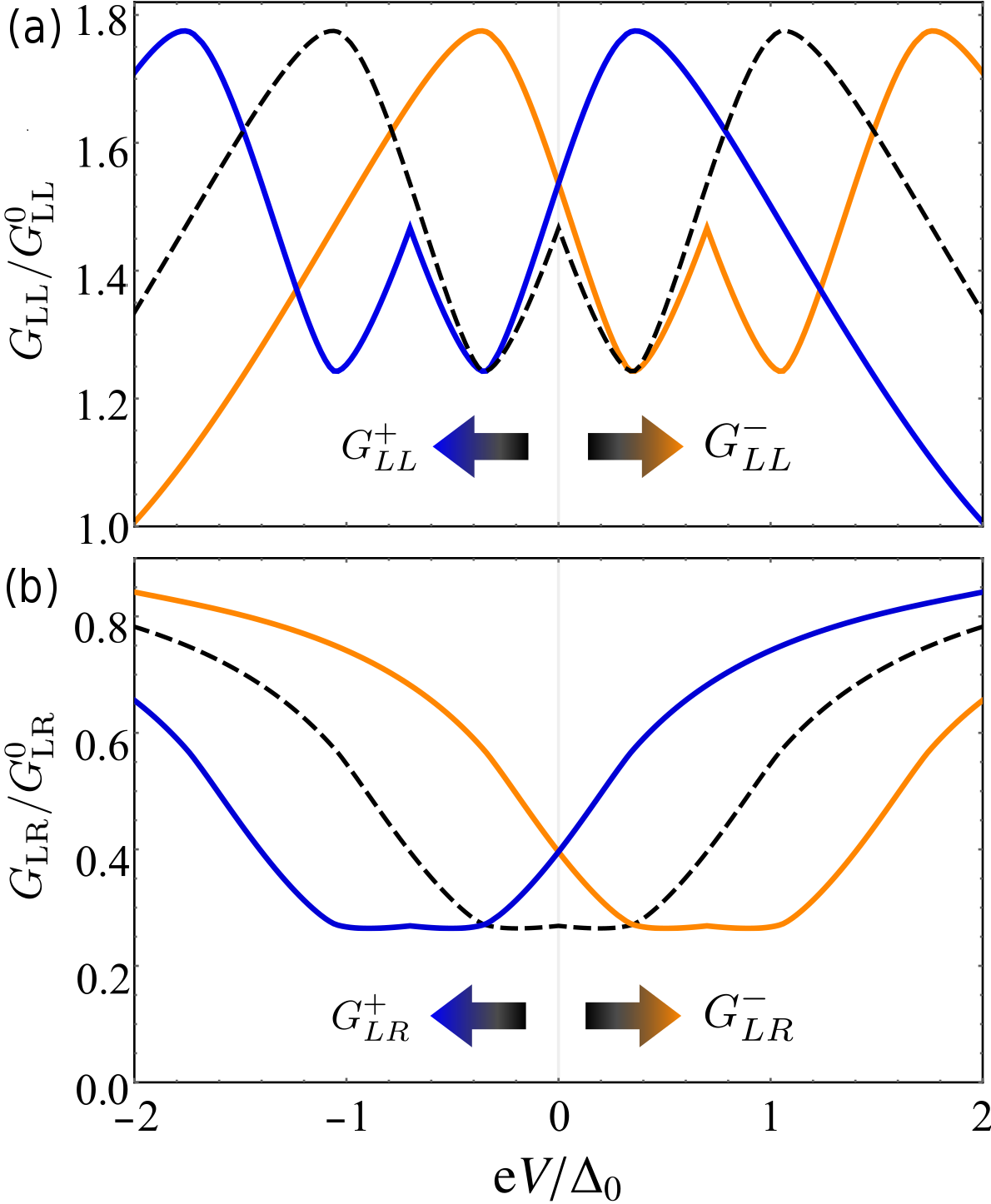}

\caption{(a) Local and (b) non\textendash local conductances $\bar{G}_{\text{L\ensuremath{\Lambda}}}^{\pm}$
as functions of the bias voltage $eV$. The black dashed curves are
for the absence of a CCP, $\chi=0$. The blue and orange solid curves
are $\bar{G}_{\text{L\ensuremath{\Lambda}}}^{+}$ and $\bar{G}_{\text{L\ensuremath{\Lambda}}}^{-}$
with $\text{\ensuremath{\Lambda}}\in\{\text{L,R\}}$ in the presence
of $\chi=0.7\Delta_{0}$, respectively. The CCP shifts the curves
of $G_{\text{L\ensuremath{\Lambda}}}^{\pm}(eV)$ in the $eV$ axis
oppositely. Here, $\alpha=\pi/2$ and other parameters are the same
as those in Fig.\ \ref{fig:conductance}.}

\label{fig:conductance-shift}
\end{figure}

For comparison, we first discuss some universal features of $\bar{G}_{\text{L\ensuremath{\Lambda}}}$
in the absence of a CCP. The results of $\bar{G}_{\text{L\ensuremath{\Lambda}}}$
are shown Fig.\ \ref{fig:conductance} as functions of the angle
$\alpha$ and bias voltage $eV$. First of all, both $\bar{G}_{\text{L\ensuremath{\Lambda}}}$
are $\pi$-periodic functions of $\alpha$ and obey a symmetry with
respect to $\alpha$, $\bar{G}_{\text{L\ensuremath{\Lambda}}}(\alpha)=\bar{G}_{\text{L\ensuremath{\Lambda}}}(\pi-\alpha)$.
This angle dependence reflects the strength of the effective superconducting
gap which scales as $\tilde{\Delta}_{0}=\Delta_{0}|\sin\alpha|$. For
$\alpha=0,\pi$, Andreev and cross-Andreev reflection processes are
completely suppressed and the same characteristics as for a Weyl NN$'$N
junction with $\Delta_{0}=0$ are observed. For $\alpha=\pi/2$ ($\tilde{\Delta}_{0}\approx\Delta_{0}$),
Andreev and crossed Andreev reflections occur with prominent probabilities
in the sub-gap region ($eV<\tilde{\Delta}_{0}$), while the normal
transport is quickly restored in the supra-gap region ($eV>\tilde{\Delta}_{0}$).
Unlike the normalized conductance in Weyl NS junctions \citep{WChen13EPL},
the normalized local conductance is no longer a constant in the sub-gap
region, as the incoming particles can tunnel from one WSM region into
the other. Instead, the conductances $\bar{G}_{\text{L\ensuremath{\Lambda}}}$
vary substantially and show oscillatory behavior which stems from
the interference effect in the junction. The local conductance $G_{\text{LL}}$
is larger than $G_{\text{LL}}^{0}$ for $\alpha\neq0,\pi$ inside
the gap due to the contribution of Andreev reflection.However, the
non-local conductance $G_{\text{LR}}$ is reduced by crossed Andreev
reflection, as indicated by Eqs.\ (\ref{eq:local-conductance}) and
(\ref{eq:nonlocal-conductance}). At zero bias voltage, $\bar{G}_{\text{LL}}$
exhibits a pronounced peak whereas $\bar{G}_{\text{LR}}$ has a flat
valley, as shown in Fig.\ \ref{fig:conductance}(c) and (d). Outside
the sub-gap region, $G_{\text{L\ensuremath{\Lambda}}}$ recover their
normal values $G_{\text{L\ensuremath{\Lambda}}}^{0}$. The oscillatory
behavior is reduced if we consider a longer junction or an angle $\alpha$
close to $0$ or $\pi$. Finally, we note that in
the absence of a CCP, both $\bar{G}_{\text{L\ensuremath{\Lambda}}}$
are even functions of $eV$. The contributions from $\mathcal{H}^{\pm}$
are identical. Thus, the measured conductances are just twice the
conductances from $\mathcal{H}^{+}$.

\begin{figure}[h]
\centering

\includegraphics[width=8.5cm]{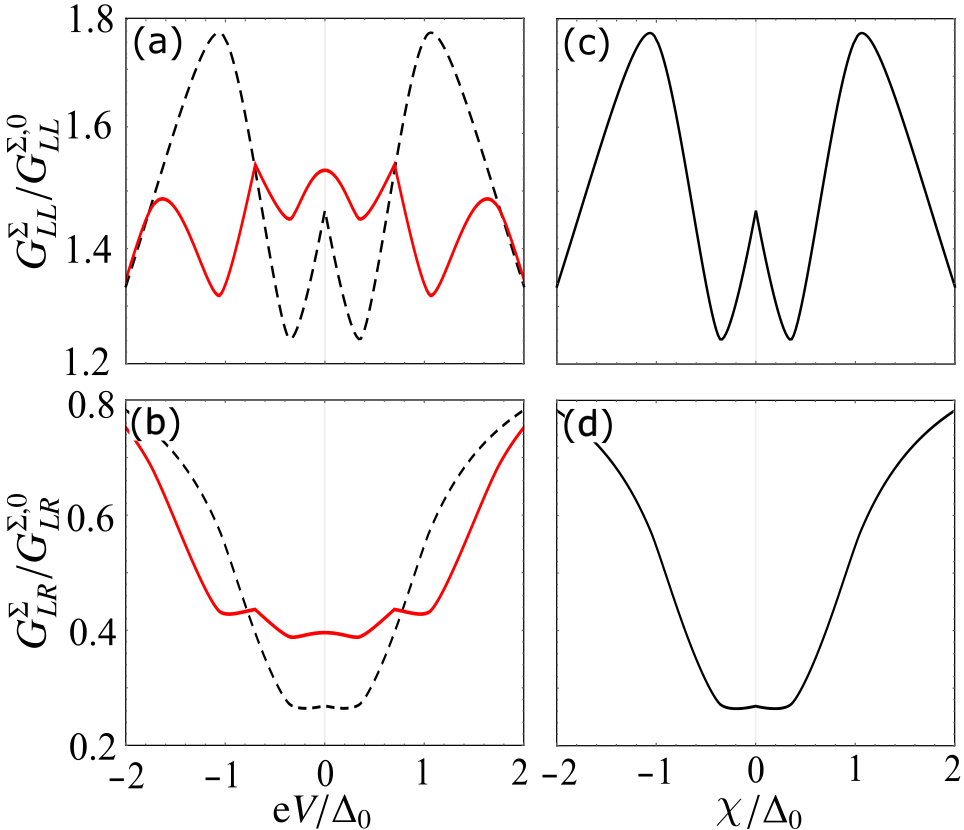}

\caption{Total (a) local and (b) non-local conductances $\bar{G}_{\text{L\ensuremath{\Lambda}}}^{\Sigma}$
as functions of the bias voltage $eV$. The black dashed and red solid
curves are for the absence and presence of a CCP $\chi=0.7\Delta_{0}$,
respectively. $\bar{G}_{\text{L\ensuremath{\Lambda}}}^{\Sigma}$ are
always even functions of $eV$. Total zero-bias (c) local and (d)
non-local conductances $\bar{G}_{\text{L\ensuremath{\Lambda}}}^{\Sigma}$
as functions of the CCP $\chi$. Here, $\alpha=\pi/2$ and other parameters
are the same as those in Fig.\ \ref{fig:conductance}.}

\label{fig:total-conductance-shift}
\end{figure}

A finite CCP enters as opposite energy shifts in the excitation energy
spectra of $\mathcal{H}^{\pm}$ and hence results in opposite shifts with respect
to the bias voltage, $eV\rightarrow eV\pm\chi$. Consequently,
the conductances $G_{\text{L\ensuremath{\Lambda}}}^{\pm}$ from the
two blocks $\mathcal{H}^{\pm}$ become different and are no longer
even functions of $eV$, see Fig.\ \ref{fig:conductance-shift}.
However, the relations between them, $G_{\text{L\ensuremath{\Lambda}}}^{-}(\chi,eV)=G_{\text{L\ensuremath{\Lambda}}}^{+}(\chi,-eV)$,
are preserved since the CCP does not break particle-hole symmetry.
Therefore, the total conductances of the system,
\begin{equation}
G_{\text{L\ensuremath{\Lambda}}}^{\Sigma}(\chi,eV)=G_{\text{L\ensuremath{\Lambda}}}^{+}(\chi,eV)+G_{\text{L\ensuremath{\Lambda}}}^{-}(\chi,\mathrm{e}V),
\end{equation}
are still even functions of $eV$. For large bias voltages $eV\gg\chi$,
the normalized conductances $\bar{G}_{\text{L\ensuremath{\Lambda}}}^{\Sigma}\equiv G_{\text{L\ensuremath{\Lambda}}}^{\Sigma}/G_{\text{L\ensuremath{\Lambda}}}^{\Sigma,0}$,
recover their values in the absence of CCP, as shown in Fig.\ \ref{fig:total-conductance-shift}(a)
and (b). However, substantial modifications emerge within the small
bias voltage window $|eV|\lesssim\chi$. In the local conductance
$\bar{G}_{\text{LL}}^{\Sigma}$, more peaks with reduced amplitudes
appear, while in the nonlocal conductance $\bar{G}_{\text{LR}}^{\Sigma}$,
the valley at zero bias voltage becomes wider and small oscillations
become possible. At zero bias voltage, $G_{\text{L\ensuremath{\Lambda}}}^{+}$
and $G_{\text{L\ensuremath{\Lambda}}}^{-}$ are always the same, as
protected by particle-hole symmetry. Thus, the total zero-bias conductances
is simply twice the value of $G_{\text{L\ensuremath{\Lambda}}}^{\pm}$.
In Fig.\ \ref{fig:total-conductance-shift}(c) and (d), we plot the
total zero-bias conductances $\bar{G}_{\text{L\ensuremath{\Lambda}}}^{\Sigma}(\chi,0)$
as functions of the CCP $\chi$. Due to the general relations between
$G_{\text{L\ensuremath{\Lambda}}}^{\pm}$, $\bar{G}_{\text{L\ensuremath{\Lambda}}}^{\Sigma}(\chi,0)$
take the exactly the same shapes as $\bar{G}_{\text{L\ensuremath{\Lambda}}}^{\Sigma}(0,eV)$.
Both $\bar{G}_{\text{L\ensuremath{\Lambda}}}^{\Sigma}(\chi,0)$ are
also even functions of $\chi$. A small $\chi$ retains almost the
same nonlocal conductance $\bar{G}_{\text{LR}}^{\Sigma}(0,eV)$ but
changes the local one $\bar{G}_{\text{LL}}^{\Sigma}(0,eV)$ quickly.
With increasing $\chi$ further, both $\bar{G}_{\text{L\ensuremath{\Lambda}}}^{\Sigma}(0,eV)$
gradually relax to unity where the superconducting effect becomes
negligible. From this point of view, the CCP plays a similar role
as the bias voltage and may be used to control the transport in the
Weyl junction.  Note that one could introduce the
CCP by applying a uniform parallel electromagnetic field \citep{Nielsen83plb,Fukushima08PRD}
or a strain deformation \citep{ZDSong16PRB,Cortijo16PRB} to the junction,
and tune the strength of the CCP via the field strength. Moreover, the conductance modification by the applied fields in the small
bias voltage regime can also serve to detect the CCP.

\section{Induced pairing amplitudes\label{sec:Induced-pairing-amplitudes}}

\begin{widetext}

\begin{figure}[h]
\centering

\includegraphics[width=17cm]{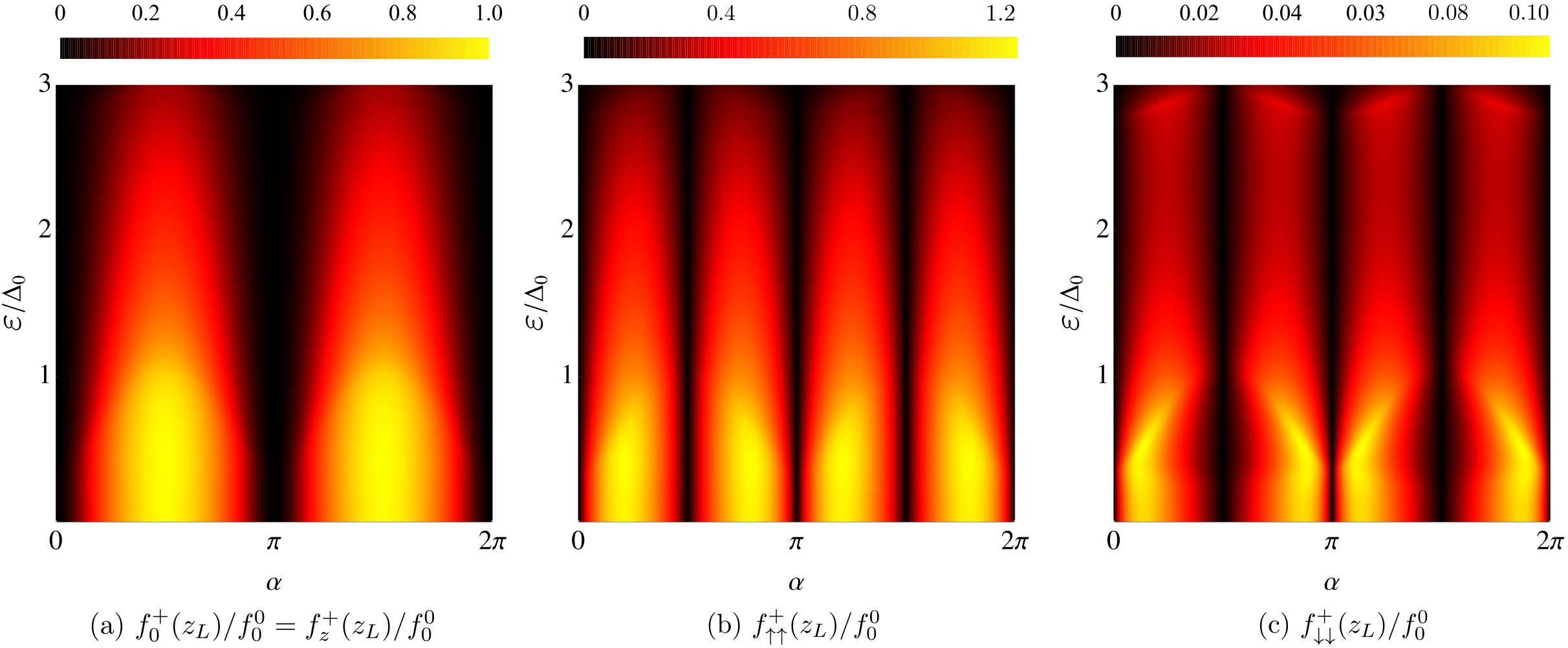}\caption{Local pairing amplitudes as functions of the angle $\alpha$ and energy
$\varepsilon$. (a) opposite-spin amplitudes $f_{0}^{+}(z_{L})/f_{0}^{0}=f_{z}^{+}(z_{L})/f_{0}^{0}$,
equal-spin amplitudes (b) $f_{\uparrow\uparrow}^{+}(z_{L})/f_{0}^{0}$
and (c) $f_{\downarrow\downarrow}^{+}(z_{L})/f_{0}^{0}$ at the left
interface $z_{L}=-L_{s}/2$. All amplitudes are normalized by $f_{0}^{0}$,
the zero-energy value of $f_{0}^{+}(z_{L})$. We choose $\mu_{S}=10^{6}\Delta_{0}$,
$\mu_{N}=10^{3}\Delta_{0}$, $\chi=0$ and $L_{s}=\xi$ for all figures.}

\label{fig:pairing}
\end{figure}

\end{widetext}

We now proceed to discuss the induced pairing amplitudes in the WSM regions. The pairing amplitudes are contained in the anomalous Green's
function (the electron-hole part of the Green's function in Nambu
space). The retarded Green's function can be constructed by combining
the scattering states\citep{Kashiwaya00RPP,McMillan68PR,Breunig18PRL}
\begin{equation}
\mathcal{G}^{R}(z,z')=\begin{cases}
\alpha_{1}\phi_{3}(z)\tilde{\phi}_{1}^{T}(z')+\alpha_{2}\phi_{3}(z)\tilde{\phi}_{2}^{T}(z')\\
+\alpha_{3}\phi_{3}(z)\tilde{\phi}_{1}^{T}(z')+\alpha_{4}\phi_{3}(z)\tilde{\phi}_{2}^{T}(z'), & z<z'\\
\\
\beta_{1}\phi_{1}(z)\tilde{\phi}_{3}^{T}(z')+\beta_{2}\phi_{1}(z)\tilde{\phi}_{4}^{T}(z')\\
+\beta_{3}\phi_{2}(z)\tilde{\phi}_{3}^{T}(z')+\beta_{4}\phi_{2}(z)\tilde{\phi}_{4}^{T}(z'), & z>z'
\end{cases}\label{eq:A11}
\end{equation}
where $\phi_{l}(z)$ with $l\in\{1,2,3,4\}$ are given by Eqs.\ (\ref{sec:Scatteringstate}),
while $\tilde{\phi}_{l}(z)$ are the scattering states of the transposed
Hamiltonian $(\mathcal{H}^{+})^{T}$, which can be obtained analytically
in a similar way. Note that $\mathcal{G}^{R}$ are also functions
of other variables such as $\varepsilon$, $\alpha$, $\chi$ and
${\bf k}_{\parallel}$. The spatial dependence is shown explicitly
in Eq.\ (\ref{eq:A11}) because it is important to determine the
coefficients $\alpha_{l}$ and $\beta_{l}$. Requiring the discontinuity
of $\mathcal{G}^{R}(z,z')$ at any position $z=z'$ across the junction,
\begin{equation}
\mathcal{G}^{R}(z,z-0^{+})-\mathcal{G}^{R}(z,z+0^{+})=-i\tau_{z}\sigma_{z},\label{eq:A12}
\end{equation}
$\alpha_{l}$ and $\beta_{l}$ are analytically derived as\begin{subequations}
\begin{align}
\alpha_{1} & =\beta_{1}=-\frac{i}{J_{e}k_{e}}\frac{c_{4}}{c_{3}c_{4}-d_{3}d_{4}},\\
\alpha_{2} & =\beta_{2}=\frac{i}{J_{h}k_{h}}\frac{d_{4}}{c_{3}c_{4}-d_{3}d_{4}},\\
\alpha_{3} & =\beta_{3}=\frac{i}{J_{e}k_{e}}\frac{d_{3}}{c_{3}c_{4}-d_{3}d_{4}},\\
\alpha_{4} & =\beta_{4}=-\frac{i}{J_{h}k_{h}}\frac{c_{3}}{c_{3}c_{4}-d_{3}d_{4}}.
\end{align}
\end{subequations}Transforming back to the original spin space, the
retarded anomalous Green's function can be written in a $2\times2$
matrix of the general form
\begin{align}
F(z,z')= & \tilde{f}_{0}^{+}(z,z')s_{0}+\tilde{f}_{\uparrow\uparrow}^{+}(z,z')(s_{x}+is_{y})\nonumber \\
 & +\tilde{f}_{\downarrow\downarrow}^{+}(z,z')(s_{x}-is_{y})+\tilde{f}_{z}^{+}(z,z')s_{z}.
\end{align}
The functions $\tilde{f}_{s}^{+}$ with $s\in\{0,z,\uparrow\uparrow,\downarrow\downarrow\}$
correspond to the pairing amplitudes of spin-singlet, opposite-spin
triplet and equal-spin triplets, respectively. In this work, we are
interested in the induced local pairing amplitudes $\tilde{f}_{s}^{+}(z)\equiv\tilde{f}_{s}^{+}(z,z'=z)$
at the interfaces and in the WSM regions. The local pairing amplitudes
in the left WSM region ($z=z'\leqslant z_{L}$) can be found explicitly
as\begin{subequations}\label{The-pairing-amplitudes}
\begin{align}
\tilde{f}_{0}^{+}(z)= & \frac{ia_{1}}{4J_{e}k_{e}}[(J_{e}J_{h}+k_{\parallel}^{2}e^{2i\theta_{k}})\sin\alpha\nonumber \\
 & -(J_{e}-J_{h})k_{\parallel}e^{i\theta_{k}}\cos\alpha]e^{-i(k_{e}+k_{h})z},\label{eq:left1}\\
\tilde{f}_{z}^{+}(z)= & -\frac{ia_{1}}{4J_{e}k_{e}}[(J_{e}J_{h}-k_{\parallel}^{2}e^{2i\theta_{k}})\sin\alpha\nonumber \\
 & -(J_{e}+J_{h})k_{\parallel}e^{i\theta_{k}}\cos\alpha]e^{-i(k_{e}+k_{h})z},\\
\tilde{f}_{\uparrow\uparrow}^{+}(z)= & -\frac{iJ_{h}a_{1}}{2J_{e}k_{e}}(J_{e}\cos\alpha-k_{\parallel}e^{i\theta_{k}}\sin\alpha)\nonumber \\
 & \times e^{-i(k_{e}+k_{h})z},\\
\tilde{f}_{\downarrow\downarrow}^{+}(z)= & -\frac{ik_{\parallel}e^{i\theta_{k}}a_{1}}{2J_{e}k_{e}}(J_{e}\sin\alpha+k_{\parallel}e^{i\theta_{k}}\cos\alpha)\nonumber \\
 & \times e^{-i(k_{e}+k_{h})z}.\label{eq:left4}
\end{align}
\end{subequations}The ones in the right WSM region ($z=z'\geqslant z_{R}$)
are related to the Andreev reflection amplitude $a_{3}$. Using $a_{3}(\alpha,\theta_{k})=a_{1}(-\alpha,-\theta_{k})$,
they can be obtained from Eqs.\ (\ref{eq:left1}-\ref{eq:left4})
by the relations\begin{subequations}
\begin{align}
\tilde{f}_{0/z}^{+}(z)|_{{\bf k}_{\parallel},\alpha} & =\mp\tilde{f}_{0/z}^{+}(-z)|_{{\bf -k}_{\parallel},-\alpha},\\
\tilde{f}_{\uparrow\uparrow/\downarrow\downarrow}^{+}(z)|_{{\bf k}_{\parallel},\alpha} & =\tilde{f}_{\downarrow\downarrow/\uparrow\uparrow}^{+}(-z)|_{{\bf -k}_{\parallel},-\alpha}.
\end{align}
\end{subequations}The local pairing amplitudes stem from Andreev
reflection at the interfaces, as indicated by their proportionality
to the Andreev reflection amplitudes $a_{1}$ or $a_{3}$.

To analyze the weights of different pairing components, we define
the averaged amplitudes as
\begin{equation}
f_{s}^{+}(z)\equiv\Big|\sum_{{\bf k}_{\parallel}}\tilde{f}_{s}^{+}(z)|_{{\bf k}_{\parallel},\alpha}\Big|,\ \ \ s\in\{0,z,\uparrow\uparrow,\downarrow\downarrow\},
\end{equation}
where the sum runs over all available modes that allow for local or
crossed Andreev reflection to happen, i.e., the processes in which
Cooper pairs are created. These quantities measure the amount of induced
Cooper pairs of spin singlet, opposite-spin triplet and two equal-spin
triplets, respectively.

Before considering the influence of the CCP, it is also instructive
to discuss some important features in the absence of a CCP. The pairing
amplitudes at the left interface $z=z_{L}$ for the block $\mathcal{H}^{+}$
are plotted in Fig.\ \ref{fig:pairing} as functions of the angle
$\alpha$ and energy $\varepsilon$. First, all the averaged amplitudes
are even functions of $\varepsilon$. With increasing $|\varepsilon|$,
$f_{0/z}^{+}(z_{L})$ decay monotonically to zero since the superconducting
effect decreases away from the Fermi energy. Second, the module of
the two opposite-spin amplitudes, namely, the spin-singlet $f_{0}^{+}(z_{L})$
and opposite-spin triplet $f_{z}^{+}(z_{L})$, are identical. This
results from the fact that for the block $\mathcal{H}^{+}$, the only
available reflected hole states moving away from the left interface
are polarized with spin up. Third, both $f_{0/z}^{+}(z_{L})$ vanish
at $\alpha=0,\pi$ whereas maximize at $\alpha=\pi/2$ and $3\pi/2$,
and they are $\pi$-periodic in $\alpha$, as shown in Fig.\ \ref{fig:pairing}(a).
This feature, similar to the conductances, stems from the $\alpha$-dependent
effective superconducting gap. In addition, we find that the opposite-spin
amplitudes at the right interface are equal to the ones at the left
interface, $f_{0/z}^{+}(z_{L})=f_{0/z}^{+}(z_{R}),$ due to a symmetry
of $\mathcal{H}^{+}$ indicated by
\begin{equation}
\sigma_{z}\mathcal{H}^{+}({\bf k}_{\parallel},k_{z},\alpha)\sigma_{z}=-\mathcal{H}^{+}({\bf k}_{\parallel},-k_{z},-\alpha).\label{eq:Symmetry}
\end{equation}
Notably, the same behavior occurs for the other block $\mathcal{H}^{-}$.

\begin{figure}
\centering

\includegraphics[width=8.5cm]{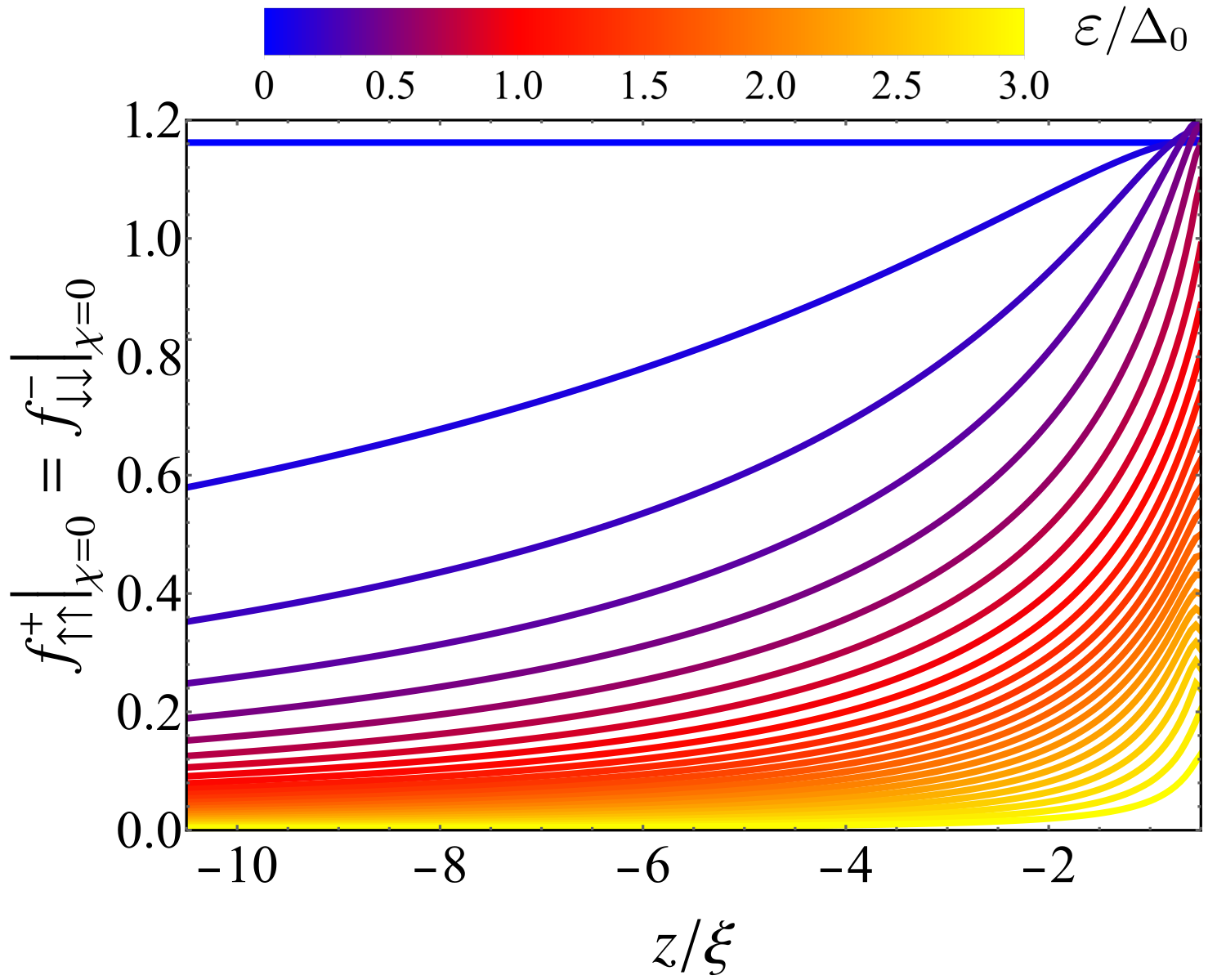}

\caption{Position dependence of the equal-spin pairing amplitudes $f_{\uparrow\uparrow}^{+}$
and $f_{\downarrow\downarrow}^{-}$ in the left WSM region for various
energies $\varepsilon$ and vanishing $\chi$. For all except the
zero energy, the amplitudes decay exponentially to zero in the WSM
region. Here, $\alpha=\pi/4$ and other parameters are the same as
those in Fig.\ \ref{fig:pairing}.}

\label{fig:pairing-imbalance-beyond-interface}
\end{figure}

More intriguing features can be found with respect to the equal-spin
amplitudes $f_{\uparrow\uparrow}^{+}$ and $f_{\downarrow\downarrow}^{+}$.
First of all, both $f_{\uparrow\uparrow/\downarrow\downarrow}^{+}(z_{L})$
vanish not only near $\alpha=0,\pi$ but also at $\alpha=\pi/2,\ 3\pi/2$,
as shown in Fig.\ \ref{fig:pairing}(b) and (c). The vanishing at
$\alpha=0,\pi$ is because of the effective pairing potential being
zero in the superconductor and no superconductivity being present
for any ${\bf k}_{\parallel}$, whereas the vanishing at $\alpha=\pi/2,\ 3\pi/2$
is due to a restoration of a $C_{4}$ symmetry with respect to $\hat{z}$
direction such that the equal-spin amplitudes from all ${\bf k}_{\parallel}$
average to zero. With increasing energy $\varepsilon$, both $f_{\uparrow\uparrow/\downarrow\downarrow}^{+}(z_{L})$
decay to zero, similar to the opposite-spin amplitudes. However, they
are no longer monotonic functions of $\varepsilon$. More interestingly,
$f_{\uparrow\uparrow}^{+}$ is quite different from $f_{\downarrow\downarrow}^{+}$
at the same interface, due to the splitting of spin degeneracy by
the strong spin-orbit coupling in the system. While $f_{\downarrow\downarrow}^{+}(z_{L})$
is at least one order in magnitude smaller than the opposite-spin
amplitudes $f_{0/z}^{+}(z_{L})$, $f_{\uparrow\uparrow}^{+}(z_{L})$
is of the same order, see Fig.\ \ref{fig:pairing}(b) and (c). Therefore,
we have a local spin polarization of Cooper pairs from $\mathcal{H}^{+}$.
Again in contrast to $f_{0/z}^{+}(z_{L})$, $f_{\uparrow\uparrow}^{+}(z_{L})$
at the left interface is different from $f_{\uparrow\uparrow}^{+}(z_{R})$
at the right interface. It is, however, equal to $f_{\downarrow\downarrow}^{+}(z_{R})$
at the right interface, and vice versa. Explicitly,
\begin{align}
 & f_{\uparrow\uparrow/\downarrow\downarrow}^{+}(z_{L})=f_{\downarrow\downarrow/\uparrow\uparrow}^{+}(z_{R}),
\end{align}
which can also be related to the symmetry in Eq.\ (\ref{eq:Symmetry})
that flips the spins.

\begin{figure}[h]
\centering

\includegraphics[width=8.5cm]{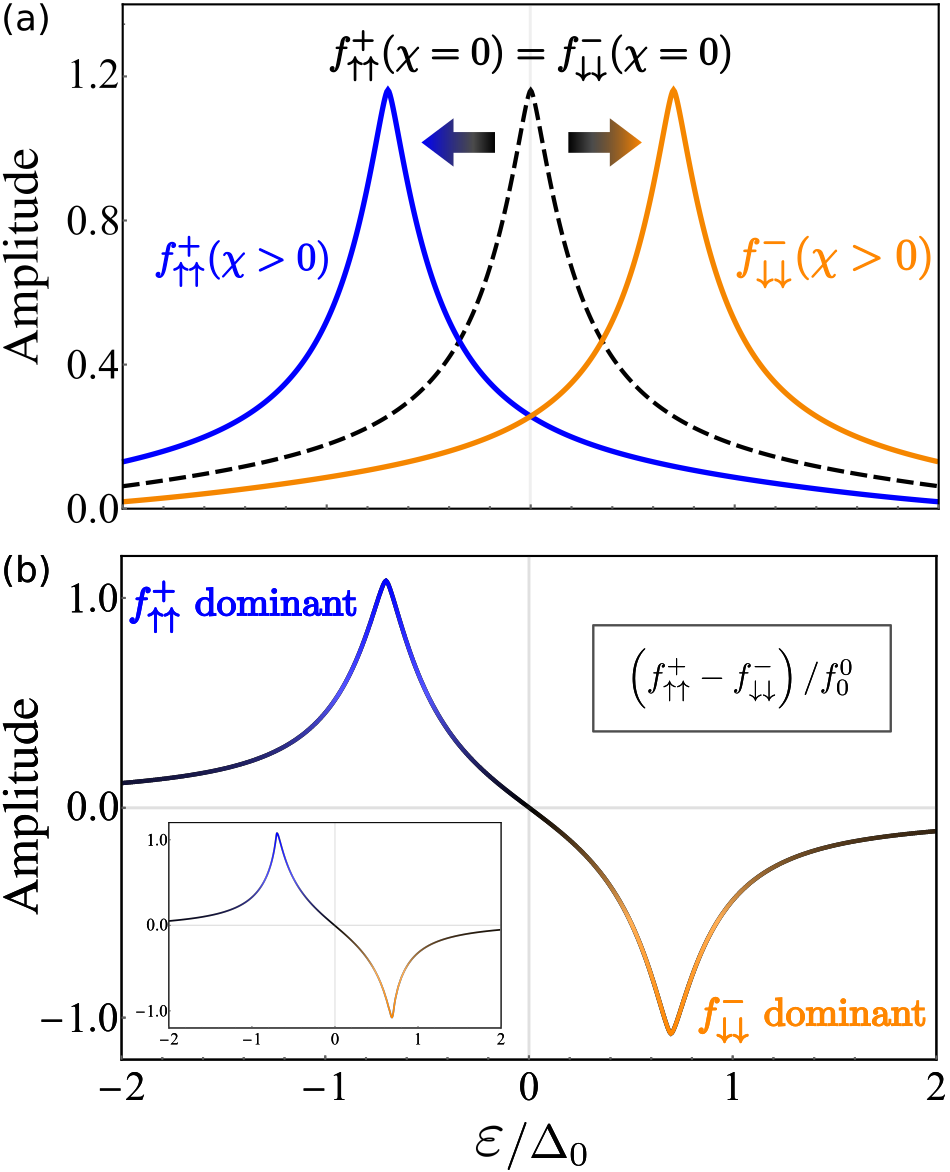}

\caption{(a) Equal-spin pairing amplitudes $f_{\uparrow\uparrow}^{+}$ and
$f_{\downarrow\downarrow}^{-}$ in the left WSM region at $z=-5\xi$
for vanishing (black and dashed line) and a finite (blue and orange
lines) CCP $\chi=0.7\varDelta_{0}$. (b) The finite CCP results in
a large net spin polarization of Cooper pairs at $\varepsilon\approx\pm\chi.$ Inset: If the CCP is present only in the leads, while absent in the superconductor, then the peak and dip are slightly skewed.
Here, $\alpha=\pi/4$ and other parameters are the same as those in
Fig.\ \ref{fig:pairing}.}

\label{fig:pairing-imbalance}
\end{figure}

We next consider the equal-spin amplitudes away from the interfaces
in the WSM regions. The equal-spin amplitudes $f_{\uparrow\uparrow}^{+}(z)$
and $f_{\downarrow\downarrow}^{-}(z)$ as functions of position $z$
inside the left WSM region for various $\varepsilon$ are presented
in Fig.\ \ref{fig:pairing-imbalance-beyond-interface}. Here, we
choose $\alpha=\pi/4$ in order to optimize the equal-spin amplitudes.
In general, $f_{\uparrow\uparrow}^{+}(z)$ and $f_{\downarrow\downarrow}^{-}(z)$
decay monotonically away from the interface into the WSM region. However,
it is interesting to find that at zero energy, $f_{\uparrow\uparrow}^{+}(z)$
and $f_{\downarrow\downarrow}^{-}(z)$ remain almost the same values
as the ones at the interface, even deep inside the WSM regions. This
can be understood from the position dependence in the amplitudes.
Take $\mathcal{H}^{+}$ for illustration. The position dependence
is contained in a phase factor
\begin{equation}
f_{\uparrow\uparrow}^{+}(z)\propto e^{-i(k_{e}+k_{h})z},
\end{equation}
according to Eqs.\ (\ref{The-pairing-amplitudes}). At zero energy,
the electron and hole wave vectors become exactly opposite, $k_{h}=-k_{e}$,
for any ${\bf k}_{\parallel}$. Then, the phase factor simply evaluates
to unity and the position dependence disappears. Therefore, for small
energies we expect considerable equal-spin pairings and a spin polarization
of Cooper pairs from the block $\mathcal{H}^{+}$ inside the WSM regions.
For the other block $\mathcal{H}^{-}$, we find, however, that the
situation is exactly opposite, i.e., $f_{\uparrow\uparrow}^{-}(z)=f_{\downarrow\downarrow}^{+}(z)$
and $f_{\downarrow\downarrow}^{-}(z)=f_{\uparrow\uparrow}^{+}(z)$.
Therefore, in total, no net spin polarization of Cooper pairs remains
in the entire system.

This scenario is dramatically changed under the influence of a finite
CCP. In Fig.\ \ref{fig:pairing-imbalance}, we plot the energy dependence
of $f_{\uparrow\uparrow}^{+}(z)$ and $f_{\downarrow\downarrow}^{-}(z)$
at $z=-5\xi$ for illustration. In the absence of a CCP, $f_{\uparrow\uparrow}^{+}$
and $f_{\downarrow\downarrow}^{-}$ are exactly the same and show
a pronounced peak at zero energy. In contrast, a CCP $\chi$ shifts
the excitation spectra of $\mathcal{H}^{\pm}$ oppositely. As a result,
$f_{\uparrow\uparrow}^{+}$ and $f_{\downarrow\downarrow}^{-}$ are
no longer degenerate but shifted oppositely by $\mp\chi$ in $\varepsilon$,
as shown in Fig.\ \ref{fig:pairing-imbalance}(a). The perfect cancellation
of the spin polarizations of Cooper pairs from the two blocks $\mathcal{H}^{\pm}$
is violated. The peak of $f_{\uparrow\uparrow}^{+}$ is moved to $\varepsilon=-\chi$,
whereas the peak of $f_{\downarrow\downarrow}^{-}$ is moved oppositely
to $\varepsilon=\chi$. In the zero and large energy limits, $f_{\uparrow\uparrow}^{+}(z)$
and $f_{\downarrow\downarrow}^{-}(z)$ still coincide and thus no
spin polarization exits. However, at energies around $\pm\chi$, $f_{\uparrow\uparrow}^{+}(z)$
and $f_{\downarrow\downarrow}^{-}(z)$ are crucially different, leading
to a large net spin polarization of Cooper pairs which is given by
the difference of the two total equal-spin amplitudes $f_{\uparrow\uparrow}(z)=f_{\uparrow\uparrow}^{+}(z)+f_{\uparrow\uparrow}^{-}(z)$
and $f_{\downarrow\downarrow}(z)=f_{\downarrow\downarrow}^{+}(z)+f_{\downarrow\downarrow}^{-}(z)$.
Note that $f_{\downarrow\downarrow}^{+}(z)$ and $f_{\uparrow\uparrow}^{-}(z)$
remain negligibly small in the left WSM region. Figure\ \ref{fig:pairing-imbalance}(b)
illustrates the spin polarization $f_{\uparrow\uparrow}(z)-f_{\downarrow\downarrow}(z)$
of Cooper pairs in $\hat{z}$ direction for a positive $\chi$. The
spin polarization is negative for positive energies whereas it is
positive for negative energies. Importantly, it exhibits a sharp peak
and a dip at $\varepsilon=\mp\chi$ as $f_{\uparrow\uparrow}^{+}(z)$
and $f_{\downarrow\downarrow}^{-}(z)$ dominate the equal-spin pairings,
respectively, which indicates the large spin polarization of Cooper
pairs. Since at $\varepsilon=\mp\chi$, the values $f_{\uparrow\uparrow}^{+}(z)$
and $f_{\downarrow\downarrow}^{-}(z)$ remain almost the same when
the position $z$ is moved more inside the WSM region, the large spin
polarization persists deep inside the WSM region. In the other WSM
region, we find a similar but opposite net spin polarization of Cooper
pairs as the CCP preserves the symmetry in Eq.\ (\ref{eq:Symmetry}).
Therefore, the hybrid junction realizes a dipole of spin-polarized
Cooper pairs.

\begin{figure}[h]
\centering

\includegraphics[width=8.5cm]{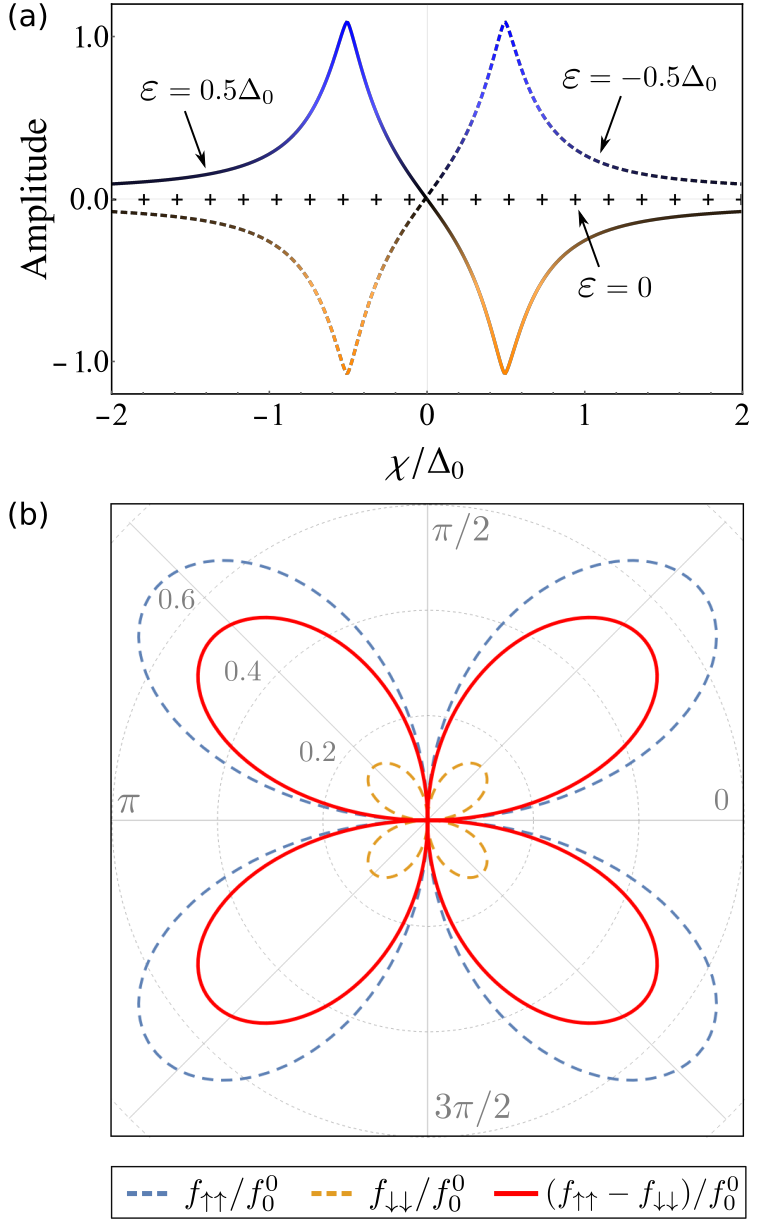}

\caption{Spin polarization $f_{\uparrow\uparrow}-f_{\downarrow\downarrow}$
at $z=-5\xi$ as a function of (a) the CCP $\chi$ for different choices
of energy $\varepsilon$ with $\alpha=\pi/4$ and (b) the angle $\alpha$
with $\varepsilon=-0.5\Delta_{0}$ and $\chi=0.7\Delta_{0}$. The
other parameters are the same as those in Fig.\ \ref{fig:pairing}.}

\label{fig:dependence}
\end{figure}

Finally, we study the dependence of the net spin
polarization of Cooper pairs on the two relevant parameters in this
work, the strength of the CCP $\chi$ and the angle $\alpha$. Choosing
the CCP to be equal everywhere in the junction, $\chi$ globally shifts
the energies $\varepsilon\to\varepsilon\pm\chi$ of $\mathcal{H^{\pm}},$
respectively. So far, we have fixed $\chi$ and varied $\omega$ in
our analysis. The behaviour of the net spin polarization should, however,
not change if $\varepsilon$ is fixed and $\chi$ is varied instead.
This is exactly what we find in Fig.\ \ref{fig:dependence}(a): choosing,
e.g., $\varepsilon=0.5\Delta_{0}$, we find the same peaks as in Fig.\ \ref{fig:pairing-imbalance}(b),
i.e., $f_{\uparrow\uparrow}^{+}(z)$ dominates at $\chi=-\varepsilon$,
while it is $f_{\downarrow\downarrow}^{-}(z)$ at $\chi=\varepsilon.$
Moreover, the zero-energy mode $\varepsilon=0$ cancels the net spin
polarization for all choices of $\chi$ as it is the case for a vanishing
CCP. This has an interesting experimental consequence, i.e., it does
not matter which quantity is fixed and which one is varied, the signature
of the net spin polarization is the same. The angle dependence inherits
mainly from those in the equal-spin amplitudes in Fig.\ \ref{fig:pairing}(b)
and (c). The spin polarization vanishes for angles $\alpha$ that
are integer multiples of $\pi/2$. When $\alpha$ deviates from these
values, the spin polarization increases quickly and reaches the maximal
value at around $\alpha_{m}\in\{\pi/4,3\pi/4,5\pi/4,7\pi/4\}$, see
Fig.\ \ref{fig:dependence}(b). The tuning of angle does, however,
not result in a qualitative change but alters merely the magnitude
in the signature. These dependencies indicate different ways to manipulate
the spin polarization of Cooper pairs.

\section{Conclusion and discussion\label{sec:Discussion-and-conclusion}}

To conclude, we have shown that the transport properties and pairing
amplitudes in Weyl NSN junctions depend on the angle between the junction
direction and the axis separating Weyl nodes. We have also found that
a CCP between Weyl nodes of opposite chirality not only modifies the
bias dependence of differential conductances, but also produces a
spin polarization of Cooper pairs in the WSM regions. The spin polarization
is opposite in the two WSM regions and can be controlled by the energy,
CCP and angle dependence.

The CCP, which can be introduced either by the chiral anomaly \citep{Nielsen83plb,Fukushima08PRD,Li16np}
or by strain deformation \citep{ZDSong16PRB,Cortijo16PRB}, is a non-equilibrium
effect. Nevertheless, we restrict ourselves in this work to a clean
system where the relaxation rate due to the inter-node scattering
is much smaller than the energy of the applied fields. In this case,
we can expect that the CCP persists for a long time and thereafter
our results are applicable.

Although the calculation presented here is based on the more realistic
assumption of a large chemical potential in the WSM compared to the
pairing potential and CCP, we note that the results of the spin polarization
of Cooper pairs and its dipolar characteristic are general and not
restricted to these assumptions. Moreover, we have also calculated
the case in which the CCP is absent in the superconductor, e.g., due
to the Meissner effect, and found that our results stay qualitatively
the same, see the inset of Fig. \ref{fig:pairing-imbalance}(b).

Recently, the time-reversal symmetry broken WSM phase has been proposed
theoretically \citep{Wan11prb,Burkov11prl,ZJWang16PRL,YJin17PRB,HYang17NJP,GQChang18PRB}
and confirmed experimentally \citep{hirschberger16nmater,kuroda17nmat,sakai18nphys}
in many realistic systems. Among these candidates, the magnetic Heusler
alloys \citep{ZJWang16PRL} and tetragonal-structured compounds \citep{YJin17PRB},
which host only a single pair of largely separated Weyl nodes, may
provide promising platforms to detect our predictions.

\section{\textup{Acknowledgments}}

We thank P. Burset, F. Dominguez, and F. Keidel for valuable discussions.
This work was supported by the DFG (SPP1666 and SFB1170 \textquotedbl ToCoTronics\textquotedbl ), the W\"urzburg-Dresden Cluster of Excellence ct.qmat, EXC2147, project-id 39085490,
and the Elitenetzwerk Bayern Graduate School on \textquotedbl Topological
Insulators\textquotedbl .


%

\end{document}